\documentclass[symmetry,article,accept,pdftex]{Definitions/mdpi}
\usepackage{amsmath,latexsym}

\firstpage{1}
\pubvolume{xx}
\issuenum{1}
\articlenumber{5}
\pubyear{2020}
\copyrightyear{2020}
\history{Received: 29 September 2020; Accepted: 12 October 2020; Published: date}

\Title{Lorentz Symmetry Group, Retardation, Intergalactic Mass Depletion and Mechanisms Leading to Galactic Rotation Curves}

\Author{Asher Yahalom $^{1,2}$\orcidA{}}

\AuthorNames{Asher Yahalom}

\address{%
$^{1}$ \quad Ariel University, Faulty of Engineering, Department of Electrical \& Electronic Engineering, Ariel 40700, Israel; asya@ariel.ac.il\\
$^{2}$ \quad Princeton University, PPPL, Princeton, NJ 08543, USA }

\abstract{The general theory of relativity (GR) is symmetric under smooth coordinate transformations, also known as diffeomorphisms.
The general coordinate transformation group has a linear subgroup denoted as the Lorentz group of symmetry, which is also maintained  in the weak field approximation to GR. The dominant operator in the weak field equation of GR is thus the d'Alembert (wave) operator, which has a retarded potential solution.
Galaxies are huge physical systems with dimensions of many tens of thousands of light years. Thus,
any change at the galactic center will be noticed at the rim only tens of thousands of years later.
Those retardation effects are neglected in the present day galactic modelling used to calculate rotational velocities of matter in the rims of the galaxy and surrounding gas. The significant differences between the predictions of Newtonian instantaneous action at a distance and observed velocities are usually explained by either assuming dark matter or by modifying the laws of gravity (MOND). In this paper,  we will show that, by taking general relativity seriously without neglecting retardation effects, one can explain the radial velocities of galactic matter in the M33 galaxy without postulating dark matter. It should be stressed that the current approach does not require that velocities $v$ are high; in fact, the vast majority of galactic bodies (stars, gas) are substantially subluminal—in other words, the ratio of $\frac{v}{c} \ll 1$. Typical velocities in galaxies are ~100 km/s, which makes this ratio $0.001$ or smaller.
However, one should consider the fact that every gravitational system, even if it is made of subluminal bodies, has a retardation distance, beyond which the retardation effect cannot be neglected.  Every natural system, such as stars and galaxies and even galactic clusters,  exchanges mass with its environment, for example, the sun loses mass through  solar wind and galaxies accrete gas from the intergalactic medium. This~means that all natural gravitational systems have a finite retardation distance. The question is thus quantitative: how large is the retardation distance? For the M33 galaxy, the velocity curve indicates that the retardation effects cannot be neglected beyond a certain distance, which was calculated to be roughly 14,000 light years; similar analysis for other galaxies of different types has shown similar results. We demonstrate, using~a detailed model, that this does not require a high velocity of gas or stars in or out of the galaxy and is perfectly consistent with the current observational knowledge of galactic and extra galactic material content and dynamics.}


\begin{document}

\newcommand{\beq} {\begin{equation}}
\newcommand{\enq} {\end{equation}}
\newcommand{\ber} {\begin {eqnarray}}
\newcommand{\enr} {\end {eqnarray}}
\newcommand{\eq} {equation}
\newcommand{\eqs} {equations }
\newcommand{\mn}  {{\mu \nu}}
\newcommand{\abp}  {{\alpha \beta}}
\newcommand{\ab}  {{\alpha \beta}}
\newcommand{\sn}  {{\sigma \nu}}
\newcommand{\rhm}  {{\rho \mu}}
\newcommand{\sr}  {{\sigma \rho}}
\newcommand{\bh}  {{\bar h}}
\newcommand{\br}  {{\bar r}}
\newcommand {\er}[1] {equation (\ref{#1}) }
\newcommand {\ern}[1] {equation (\ref{#1})}
\newcommand {\Ern}[1] {Equation (\ref{#1})}
\newcommand{\hdz}  {\frac{1}{2} \Delta z}

\section {Introduction}

The general theory of relativity (GR) is known to be symmetric under general coordinate modifications (diffeomorphism).
This group of general coordinate transformations has a Lorentz subgroup, which is maintained even in the weak field approximation. This
is manifested through the field equations containing the d'Alembert (wave) operator, which has a retarded potential~solution.

From an observational point of view, it is well known that GR is verified by many observations.
However, at~the present time, the~standard Newton–Einstein gravitational theory stands at something of a crossroads. It simultaneously has much in its favor observationally, while, at~the same time, it has some very disquieting challenges. The~observational successes that it has achieved in both astrophysics and cosmology have to be tempered by the fact that the theory needs to appeal to two as yet unconfirmed ingredients, dark matter and dark energy, in~order to achieve these successes. The~dark matter problem has not only been with us since the 1920s and 1930s (when it was initially known as the missing mass problem), but~it has also become more and more severe as more and more dark matter has had to be introduced on larger and larger distance scales as new data have come online. Moreover, extensive—now 40-year—underground and accelerator searches have failed to find any of it or establish its existence. The~dark matter situation has become even more dire in the last few years as the Large Hadron Collider has failed to find any super symmetric particles, not only of the community's preferred form of dark matter, but~also the form of it that is required in string theory, a~theory that attempts to provide a quantized version of Newton–Einstein~gravity.

While things may still eventually work out in favor of the standard dark matter paradigm, the~situation is disturbing enough to warrant consideration of the possibility that the standard paradigm might at least need to be modified in some way if not outright replaced. The~present proposal sets out to seek such a modification. Unlike other approaches such as Milgrom's MOND, Mannheim's Conformal Gravity or Moffat's MOG, the~present approach is, in~a sense, the~minimalist one adhering strictly to the basic scientific principle dictated by Occam's razor.
It seeks to replace dark matter by effects within standard General Relativity~itself.

 In 1933, Fritz Zwicky noticed that the velocities of a set of Galaxies within the Comma Cluster are much higher than those predicted by the virial calculation that assumes Newtonian theory~\cite{zwicky}.  He~calculated that the amount of matter required to
  account for the velocities could be 400 times greater with respect to that of visible matter, which led to suggesting  dark
   matter throughout the entire cluster.  Volders, in~1959, indicated  that  stars in the outer rims of the nearby spiral galaxy M33
   do not move ``correctly''~\cite{volders}. It is the result of the virial theorem coupled with Newtonian Gravity  which implies that $MG/r \sim M v^2$, that is to say, the~expected rotation curve should at some point decrease as $1/\sqrt{r}$.
   This was well established during the seventies when Rubin and Ford~\cite{rubin1,rubin2} demonstrated that, for~a large sample of spiral galaxies, this behavior can be considered a general feature: velocities at the our rim of the galaxies do not decrease—rather, in~a general case, they attain a plateau at some velocity, which is different for each galaxy.
    In what follows, we will show that such effects can be deduced from GR if retardation effects are not neglected. The~derivation of the retardation force described in previous publications~\cite{YaRe1,ge,YaRe2} is repeated for completeness. However, a~fit of the theory to the M33 rotation galaxy is given for the first time.
     It should be stressed that the current approach does not require that velocities $v$ are high; in fact, the~vast majority of galactic bodies (stars, gas) are substantially subluminal—in other words, the~ratio of $\frac{v}{c} \ll 1$. Typical velocities in galaxies are ~100~km/s, which makes this ratio $0.001$ or smaller.
However, one should consider the fact that every gravitational system, even if it is made of subluminal bodies, has a retardation distance, beyond~which the retardation effect cannot be neglected.  Every natural system, such as stars and galaxies and even galactic clusters,  exchanges mass with its environment. For~example, the~sun loses mass through solar wind and galaxies accrete gas from the intergalactic medium. This means that all natural gravitational systems have a finite retardation distance. The~question is thus quantitative: how large is the retardation distance? The~change in the mass of the sun is quite small and thus the retardation distance of the solar system is quite large, allowing us to neglect retardation effects within the solar system. However, for~the M33 galaxy, the~velocity curve indicates that the retardation effects cannot be neglected beyond a certain distance smaller than 14,000 light years; similar analyses for other galaxies of different types have shown similar results~\cite{Wagman,Wagman2}. We~demonstrate, in~Section~\ref{dynmodel}, using a detailed model, that this does not require a high velocity of gas or stars in or out of the galaxy and is perfectly consistent with the current observational knowledge of galactic and extra galactic material content and~dynamics.

\section {General Relativity}

The general theory of relativity is based on two fundamental equations, the~first being Einstein equations~\cite{Narlikar,Weinberg,MTW,Edd}:
\beq
G_\mn = -\frac{8 \pi G}{c^4} T_\mn
\label{ein}
\enq
$G_\mn$ stands for the Einstein tensor, $T_\mn$ indicates the
stress–energy tensor, $G \simeq 6.67 \ 10^{-11} \ \text{m}^3 \text{kg}^{-1} \text{s}^{-2}$ is the universal gravitational constant and $c \simeq 3 \ 10^8  \ \text{m s}^{-1}$ indicates the velocity of light in the absence of matter (Greek letters are indices in the range $0-3$).
The second fundamental equation that GR is based on is the geodesic equation:
\beq
\frac{d^2 x^\alpha}{ds^2}+\Gamma^\alpha_\mn \frac{d x^\mu}{ds} \frac{d x^\nu}{ds} = \frac{d u^\alpha}{ds}+\Gamma^\alpha_\mn u^\mu u^\nu = 0
\label{geo}
\enq
$x^\alpha (s)$ are the coordinates of the particle in spacetime, $s$ is a typical parameter along the trajectory that for massive particles is chosen to be the length of the trajectory, $u^\mu = \frac{d x^\mu }{d s}$  is the $\mu$-th component of the 4-velocity of a massive particle moving along the geodesic trajectory $s$ (increment of $x$ per unit length $s$) and $\Gamma^\alpha_\mn$ is the affine connection (Einstein summation convention is assumed). The~stress–energy tensor of matter is usually taken in the form:
\beq
T_\mn = (p+\rho c^2) u_\mu  u_\nu - p \ g_\mn
\label{fltens}
\enq

In the above, $p$ is the pressure and $\rho$ is the {\bf mass} density. We remind the reader that lowering and raising
indices is done through the metric $g_\mn$ and  inverse metric $g^\mn$, such that $u_\mu= g_\mn u^\nu$. The~same metric serves to calculate $s$:
\beq
ds^2 = g_\mn dx^\mu  dx^\nu,
\label{intervale}
\enq
 and the affine connection:
\beq
\Gamma^\alpha_\mn = \frac{1}{2} g^\ab \left(g_{\beta \mu, \nu} + g_{\beta \nu, \mu} - g_{\mn, \beta}\right), \qquad
 g_{\beta \mu, \nu} \equiv \frac{\partial g_{\beta \mu}}{\partial x^\nu}
\label{affine}
\enq

The affine connection serves to calculate the Riemann and Ricci tensors and the curvature scalar:
\beq
R^\mu_{\nu \ab} = \Gamma^\mu_{\nu \alpha,\beta} - \Gamma^\mu_{\nu \beta,\alpha} + \Gamma^\sigma_{\nu \alpha} \Gamma^\mu_{\sigma \beta}
- \Gamma^\sigma_{\nu \beta} \Gamma^\mu_{\sigma \alpha}, \quad R_{\ab}= R^\mu_{\ab \mu}, \quad R= g^\ab R_{\ab}
\label{RieRicci}
\enq
which, in~turn, serves to calculate the Einstein tensor:
\beq
G_{\ab}= R_{\ab} - \frac{1}{2} g_\ab R.
\label{Eint}
\enq

Hence, the~given matter distribution determines the metric through Equation~(\ref{ein}) and the metric determines the geodesic trajectories through Equation~(\ref{geo}). Those equations are well known to be symmetric under smooth coordinate transformations (diffeomorphism).
\beq
x'_{\alpha}= x'_{\alpha} (x_\mu).
\label{Gct}
\enq

\section {Linear Approximation of GR}

Only in extreme cases of compact objects (black holes and neutron stars) and the primordial reality or the very early universe does one need not consider the solution of the full non-linear Einstein Equation~\cite{YaRe1}. In~typical cases of astronomical interest (the galactic case included) one can use a linear approximation to those equations around the flat Lorentz metric $\eta_{\mn}$ such that (Private communication with the late Professor Donald Lynden-Bell):
 \beq
 g_{\mn} = \eta_{\mn} + h_{\mn}, \quad \eta_{\mn} \equiv \ {\rm diag } \ (1,-1,-1,-1),
 \qquad
 |h_{\mn}|\ll 1
 \label{lg}
 \enq

 One {\bf then} defines the quantity:
 \beq
 \bar h_\mn \equiv h_\mn -  \frac{1}{2} \eta_\mn h, \quad h = \eta^{\mn} h_{\mn},
 \label{bh}
 \enq
 $\bar h_\mn = h_\mn $ for non diagonal terms. For~diagonal terms:
 \beq
 \bar h = - h \Rightarrow  h_\mn = \bar h_\mn -  \frac{1}{2} \eta_\mn \bar h .
 \label{bh2}
 \enq

  The general coordinate transformation symmetry of Equation~(\ref{Gct}) has a subgroup  of infinitesimal transformations which are manifested in the gauge freedom of $h_\mn$ in the weak field approximation. It~can be shown (\cite{Narlikar} page 75, exercise 37, see also~\cite{Edd,Weinberg,MTW}) that one can  choose a gauge such that the Einstein equations are:
 \beq
\Box \bh_{\mn} \equiv \bh_{\mn, \alpha}{}^{\alpha}=-\frac{16 \pi G}{c^4} T_\mn , \qquad \bh_{\mu \alpha,}{}^{\alpha}=0.
\label{lineq1}
\enq

The d'Alembert operator $\Box$ is clearly invariant under the Lorentz symmetry group (another subgroup of the general coordinate transformation symmetry described by Equation (\ref{Gct})), of~which the Newtonian Laplace operator
$\vec \nabla^2$ is not, but~this comes with the price that "action at a distance" solutions are forbidden and only retarded solutions are
allowed. The~$T_\mn$ stress energy tensor should be calculated at the appropriate frame and thus for the rotation curve frame of reference
for galaxies, matter is approximately at~rest.

\Ern{lineq1} can always be integrated to take the form~\cite{Jackson} (For reasons why the symmetry between space and time
is broken, see~\cite{Yahalom,Yahalomb}):
 \ber
& & \bh_{\mn}(\vec x, t) = -\frac{4 G}{c^4} \int \frac{T_\mn (\vec x', t-\frac{R}{c})}{R} d^3 x',
\nonumber \\
 t &\equiv& \frac{x^0}{c}, \quad \vec x \equiv x^a \quad a,b \in [1,2,3], \nonumber \\
  \vec R &\equiv& \vec x - \vec x', \quad R= |\vec R |.
\label{bhint}
\enr

The factor before the integral is small: $\frac{4 G}{c^4} \simeq 3.3 \times 10^{-44}$; hence, in~the above calculation
one can take $T_\mn$, which is zero order in $h_\abp$.
Let us now calculate the affine connection in the linear approximation:
\beq
\Gamma^\alpha_\mn = \frac{1}{2} \eta^\abp \left(h_{\beta \mu, \nu} + h_{\beta \nu, \mu} - h_{\mn, \beta}\right).
\label{affinel}
\enq

The affine connection has only first order terms in $h_\abp$; hence, to~the first order
$\Gamma^\alpha_\mn u^\mu u^\nu$ appearing in the geodesic,  $u^\mu u^\nu$ is of the zeroth order. In~the zeroth order:
\beq
u^0=\frac{1}{\sqrt{1-\frac{v^2}{c^2}}}, u^a = \vec u =\frac{\frac{\vec v}{c}}{\sqrt{1-\frac{v^2}{c^2}}} ,
\vec v \equiv  \frac{d \vec x}{d t}, \quad v= |\vec v|.
\label{uz}
\enq

For non relativistic velocities:
\beq
u^0 \simeq 1,  \qquad \vec u \simeq \frac{\vec v}{c} , \qquad u^a \ll u^0   \qquad {\rm for} \quad v \ll c.
\label{uzslo}
\enq

Hence, we will not be considering  a post-Newtonian approximation in this paper, in~which matter travels at nearly relativistic speeds, but~
we will be considering the retardation effects and finite propagation speed of the gravitational field. We underline that
taking $\frac{v}{c} << 1$ is not the same as taking $\frac{R}{c} << 1$ (with $R$ being the typical size of a galaxy) since:
\beq
\frac{R}{c} = \frac{v}{c} \frac{R}{v}
\label{retarba}
\enq
and, in~galaxies, $\frac{R}{v}$ is a very big number ($\frac{R}{v} \simeq 10^{15}$ seconds); thus, $\frac{v}{c}$ can be neglected but
not $\frac{R}{c}$, in~which $\frac{R}{c} \simeq 10^{12}$ seconds.
By inserting Equations~(\ref{affinel}) and (\ref{uzslo}) in the geodesic equation, we arrive at the approximate form:
\beq
\frac{d v^a}{dt}\simeq - c^2 \Gamma^a_{00} = - c^2 \left( h^a_{0,0} - \frac{1}{2} h_{00,}{}^a \right)
\label{geol}
\enq

Let us now look at Equation~(\ref{fltens}). We~assume $\rho c^2 \gg p$ and, taking into account Equation~(\ref{uzslo}), we~arrive at $T_{00} = \rho c^2 $, while  other tensor components are significantly smaller. Thus, $\bar h_{00}$ is significantly larger than other components of $\bar h_\mn$. One should notice that it is not possible to deduce from the magnitudes of quantities that such a difference
exists between their derivatives. In~fact, by~the gauge condition in Equation~(\ref{lineq1}):
\beq
\bar h_{\alpha 0,}{}^0=-\bar h_{\alpha a,}{}^a \qquad \Rightarrow
\bar h_{00,}{}^0=-\bar h_{0 a,}{}^a, \quad \bar h_{b0,}{}^0=-\bar h_{b a,}{}^a.
\label{gaugeim}
\enq

Thus, the~zeroth derivative of $\bar h_{00}$ (which contains a $\frac{1}{c}$) is the same order as the spatial derivative
of $\bar h_{0a}$ and the zeroth derivative of $\bar h_{0a}$ (which appears in Equation~(\ref{geol})) is the same order
as the spatial derivative of $\bar h_{ab}$. However, it we can compare spatial derivatives of $\bar h_{00}$ and $\bar h_{ab}$
and conclude that the former is larger than the later. Using Equation~(\ref{bh2}) and taking into account the above consideration,
we may write Equation~(\ref{geol}) as:
\beq
\frac{d v^a}{dt}\simeq \frac{c^2}{4} \bar h_{00,}{}^a \Rightarrow \frac{d \vec v}{dt} = - \vec \nabla \phi = \vec F,
\qquad \phi \equiv \frac{c^2}{4} \bar h_{00}
\label{geol2}
\enq

Thus, $\phi$ is the gravitational potential of the motion which can be calculated using Equation~(\ref{bhint}):
\ber
\phi &=& \frac{c^2}{4} \bar h_{00}
= -\frac{ G}{c^2} \int \frac{T_{00} (\vec x', t-\frac{R}{c})}{R} d^3 x'
\nonumber \\
&=& -G \int \frac{\rho (\vec x', t-\frac{R}{c})}{R} d^3 x'
\label{phi}
\enr
and $\vec F$ is the force per unit mass. If~the mass density $\rho$ is static, we are in the realm of the Newtonian instantaneous action at a distance. We point out that it is unlikely that $\rho$ is static, as~a galaxy will obtain mass from the intergalactic~medium.

\section {Beyond the Newtonian Approximation}

The retardation time $\frac{R}{c}$ may be a few tens of thousands of years, but~can be considered short in comparison
to the time taken for the galactic density to change significantly. Thus, we can write a Taylor series for the density:
\beq
\rho (\vec x', t-\frac{R}{c})=\sum_{n=0}^{\infty} \frac{1}{n!} \rho^{(n)} (\vec x', t) (-\frac{R}{c})^n,
\qquad \rho^{(n)}\equiv \frac{\partial^n \rho}{\partial t^n}.
\label{rhotay}
\enq

By inserting Equations~(\ref{rhotay}) into Equation~(\ref{phi}) and keeping the first three terms, we will obtain:
\ber
\phi &=& -G \int \frac{\rho (\vec x', t)}{R} d^3 x' +  \frac{G}{c}\int \rho^{(1)} (\vec x', t) d^3 x'
\nonumber \\
&-& \frac{G}{2 c^2}\int R \rho^{(2)} (\vec x', t) d^3 x'
\label{phir}
\enr

The Newtonian potential is the first term, the~second term has null contribution, and~the third term is the lower order correction to the Newtonian theory:
\beq
 \phi_r = - \frac{G}{2 c^2} \int  R \rho^{(2)} (\vec x', t) d^3 x'
\label{phir2}
\enq

The total force per unit mass:
\ber
\vec F &=& \vec F_N + \vec F_r
\nonumber \\
 \vec F_N &=& - \vec \nabla \phi_N =  -G \int \frac{\rho (\vec x',t)}{R^2} \hat R d^3 x', \qquad \hat R \equiv \frac{\vec R}{R}
\nonumber \\
 \vec F_r &\equiv& - \vec \nabla \phi_r =  \frac{G}{2 c^2} \int  \rho^{(2)} (\vec x', t) \hat R d^3 x'
\label{Fr}
\enr

While the Newtonian force $\vec F_N$ is always attractive, we notice that the retardation force $\vec F_r$ can be
 attractive or repulsive. The~Newtonian force decreases as $\frac{1}{R^2}$,  but~the retardation force is independent of distance as long as the Taylor approximation given in Equation~(\ref{rhotay}) is valid. Below~a certain distance, the~Newtonian force will be dominant, but~for higher distances the retardation force becomes larger. Newtonian force should be neglected for distances significantly larger than the retardation distance, defined as:
\beq
 R \gg R_r \equiv c \Delta t
\label{Rr}
\enq
$\Delta t$ is the typical duration in which the density $\rho$ changes; this will be defined more accurately later. On~the other hand, for~$R\ll R_r$, the~retardation effect can be neglected and only Newtonian forces should be considered; this is probably the situation in the solar system.  It should be stressed that the existence of $R_r$ does not require that velocities $v$ are high; in fact, the~vast majority of galactic bodies (stars, gas) are substantially subluminal. In~other words, the~ratio of $\frac{v}{c} \ll 1$. Typical velocities in galaxies are $100$ km/s, which makes this ratio $0.001$ or smaller.
However, one should consider the fact that every gravitational system, even if it is made of subluminal bodies, has a retardation distance, beyond~which the retardation effect cannot be neglected.  Every natural system, such as stars and galaxies and even galactic clusters,  exchanges mass with its environment. For~example, the~sun loses mass through solar wind and galaxies accrete gas from the intergalactic medium. This means that all natural gravitational systems have a finite retardation distance. The~question is thus quantitative: how large is the retardation distance? This question will be further addressed in Section~\ref{RoCu}.
For large distances, $r=|\vec x|\rightarrow\infty$, such that $\hat R \simeq \frac{\vec x}{|\vec x|} \equiv \hat r$; thus, we obtain:
\beq
\vec F_r =  \frac{G}{2 c^2} \hat r \int  \rho^{(2)} (\vec x', t)  d^3 x' =  \frac{G}{2 c^2} \hat r \ddot{M}, \qquad
\ddot{M} \equiv \frac{d^2 M}{dt^2}.
\label{Fr2}
\enq

As the galaxy attracts intergalactic gas, its mass becomes larger and therefore $\dot{M}>0$;
however, as~the intergalactic gas is depleted, the~rate at which the mass is accumulated must decrease and therefore $\ddot{M}<0$. Thus, in~the
galactic case:
\beq
\vec F_r =  - \frac{G}{2 c^2}  |\ddot{M}| \hat r
\label{Fr3}
\enq
and the retardation force is~attractive.

\section{Rotation~Curves}
\label{RoCu}

 To calculate the non-asymptotic rotation curve, we note that the square of the azimuthal (orbital) velocity, $v_\theta$ divided by $\br$, is proportional to the derivative of the gravitational potential by $\br$, as~shown in  Equation~(\ref{Eulersr2}).  We use the cylindrical coordinates $\br,\theta,z$, in~which $z=0$ is the galactic plane. The gravitational potential will be evaluated using Equation~(\ref{phir}),  which is composed of both the Newtonian $\phi_N$ and retardation $\phi_r$ parts, such that:
\beq
 \frac{v_\theta^2}{\br} =  \frac{\partial \phi }{\partial \br} =\frac{\partial \phi_N }{\partial \br} + \frac{\partial \phi_r }{\partial \br} ,
\label{Eulersr3}
\enq

To do this, we first need a density profile describing the mass distribution in and near the galactic~plane.

\subsection{General~Considerations}

We assume a density distribution of the form:
\beq
\rho(\vec x,t)=\rho_a (\vec x) + \rho_b (\vec x)g'(t)
 \label{rhog}
\enq

Although the density profiles $\rho_a$ and $\rho_b$ need not have similar forms, we will assume, for~simplicity, that they do. Hence,
$\rho_a=\rho_b$ and, by~defining $g(t) = 1 + g'(t)$, we obtain a density distribution of the~form:
\beq
\rho(\vec x,t) = g(t) \rho_a (\vec x)
\label{rhotd};
\enq
hence,
\beq
M(t) = \int d^3 x' \rho(\vec x',t) = g(t) \int d^3 x' \rho_a (\vec x')
\label{M}
\enq
and
\beq
\ddot{M}(t) =  \ddot{g}(t) \int d^3 x' \rho_a (\vec x'),
\label{ddM2}
\enq
which leads to the result:
\beq
  |\ddot{g}(t)| = \frac{|\ddot{M}(t)|}{M(t)} g(t)  = \frac{c^2}{R_r^2} g(t);
\label{ddg}
\enq
here, we define:
\beq
t_r\equiv \sqrt{\frac{M}{|\ddot{M}|}}, \qquad \Delta t = t_r \Rightarrow R_r \equiv c t_r.
\label{deltat}
\enq

Inserting Equations~(\ref{rhotd}) and (\ref{ddg}) into Equation~(\ref{phir2}) leads to the following form of the retardation~potential:
\beq
 \phi_r = - \frac{G}{2 c^2} \ddot{g}(t) \int  R \rho_a (\vec x') d^3 x' =  \frac{G}{2 R_r^2} \int  R \rho (\vec x',t) d^3 x'.
\label{phir3}
\enq

Now, we introduce the dimensionless quantities:
\beq
\tilde{\rho} \equiv \frac{\rho}{\rho_c}, \qquad \tilde{x} \equiv \frac{\vec x}{R_s}
\label{dless}
\enq
in terms of a typical density $\rho_c$ and a typical scale $R_s$. From~now on, we will consider all time-dependent quantities as being given at the time the galaxy is being observed. This has nothing to do with the present situation in a certain galactic system that may be millions or billions year in the future with respect to the data available to us,  which of course is also the result of retardation. The~dependence of the temporal variable $t$ will thus be omitted. Using Equations~(\ref{dless})  and (\ref{M}) will take the form:
\beq
M = \int d^3 x' \rho(\vec x') = \rho_c R_s^3 \int d^3 \tilde x' \tilde \rho(\tilde x')
\label{M2}
\enq

We now define the dimensionless constant:
\beq
\Lambda = \int d^3 \tilde x' \tilde \rho(\tilde x')
\label{lam2}
\enq

In terms of which:
\beq
M =  \Lambda \rho_c R_s^3
\label{M3}
\enq

Hence, $\phi_r$ can be written as:
\beq
 \phi_r =  \frac{G M}{2 R_r^2} r \ \chi, \qquad \chi \equiv  \frac{1}{\Lambda} \int  \frac{R}{r} \tilde \rho (\tilde x') d^3 \tilde x'
\label{phir4}
\enq
$\chi$ is a dimensionless function that satisfies:
\beq
\lim_{r->\infty} \chi  = 1
\label{chi}
\enq

Similarly, one can write the Newtonian potential as:
\beq
 \phi_N = - \frac{G M}{r} \ \psi, \qquad \psi \equiv  \frac{1}{\Lambda} \int  \frac{r}{R} \tilde \rho (\tilde x') d^3 \tilde x'
\label{phiN4}
\enq
$\psi$ is a dimensionless function that also satisfies:
\beq
\lim_{r->\infty} \psi  = 1
\label{psi}
\enq

\subsection{M33 Density~Profile}

\begin{figure}[H]
\centering
\includegraphics[width=\columnwidth]{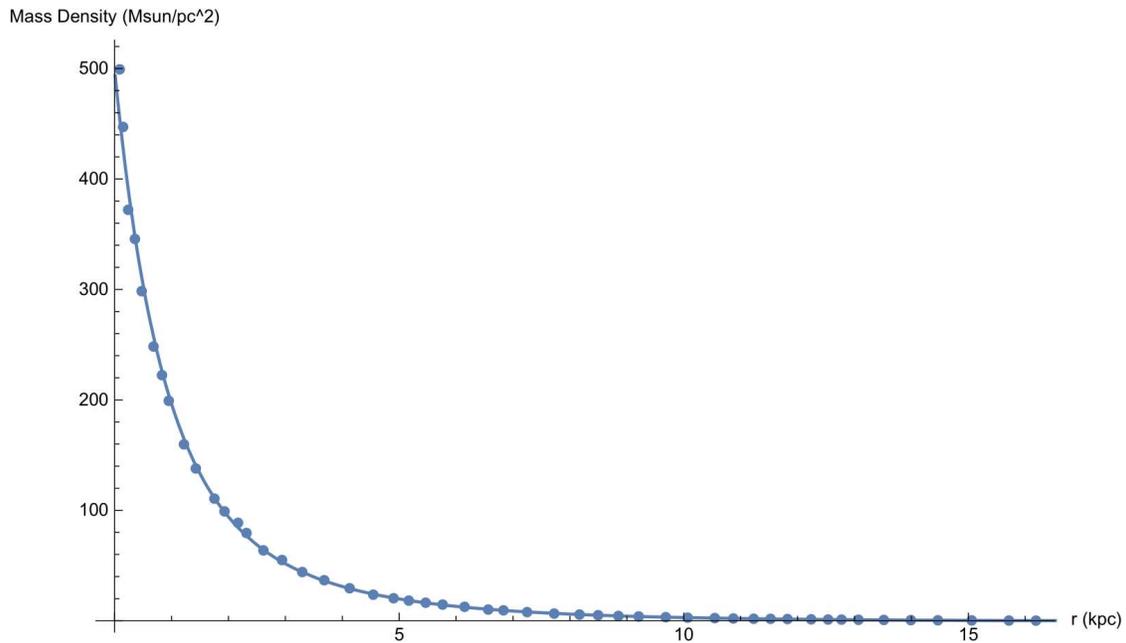}
 \caption{The M33 density radial distribution, the~dots are data points~\cite{Corbelli2} the solid line is a fit Equation~(\ref{kp}).}
 \label{radialdis}
\end{figure}

In order to calculate the retardation and Newtonian gravitational potentials and hence the rotation curve through Equation~(\ref{Eulersr2}), we must know the density distribution in the galactic plane. This is usually done as follows: for the radial distribution, we will use the luminosity distribution and assume a proportionality constant known as the mass to light ratio. Such radial density distribution is given in Figure~\ref{radialdis}, which is based on the work of Corbelli~\cite{Corbelli2} (see also Rega and Vogel~\cite{REV}) and its logarithm is fitted to a polynomial of order five, as~described in Equation~(\ref{kp}):
\ber
\rho(\br) &\sim& e^{kp (\br)},
\nonumber \\
 kp (\br) &=& 6.21207 - 1.05618 \br + 0.137599 \br^2 - 0.0149017 \br^3
 \nonumber \\
 &+&  0.000865154 \br^4 - 0.0000209513 \br^5
\label{kp}
\enr

\begin{figure}[H]
\centering
\includegraphics[width=0.58\columnwidth]{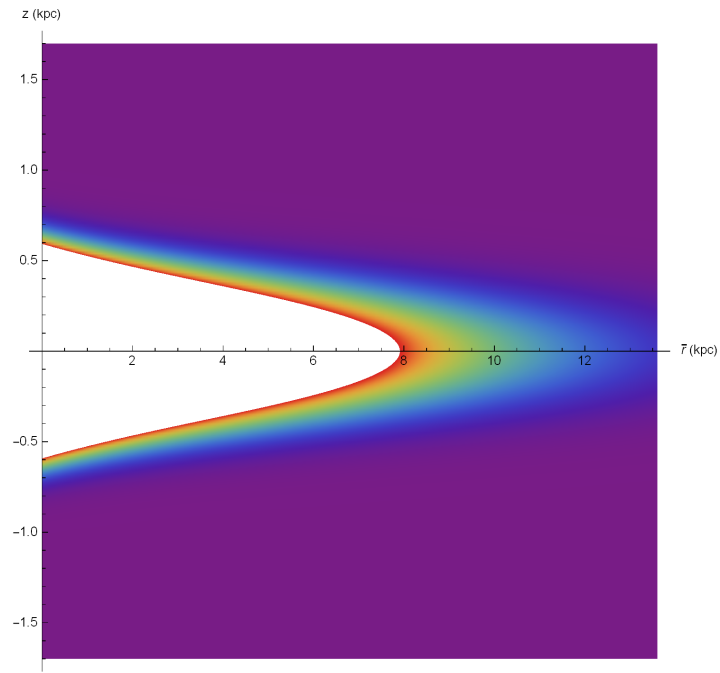}
 \caption{The M33 three dimensional mass density distribution~model.}
 \label{3Ddis}
\end{figure}

For the $z$ direction distribution orthogonal to the galactic plane, we will assume a Gaussian profile with a typical width of $\sigma = 0.2$ kpc.
The mass distribution can be written as:
\beq
\rho(\br,z) = \rho_c  e^{kp (\br)} e^{-\frac{z^2}{\sigma^2}},
\label{dens}
\enq
in which we assume cylindrical symmetry. The~three-dimensional galactic mass distribution is depicted in Figure~\ref{3Ddis}.
\unskip

\newpage
\subsection{M33 Rotation~Curve}

To obtain the rotation curve we need to evaluate the functions $\psi(\br)$ and $\chi(\br)$, which describe the non-trivial contribution of the  Newtonian and retardation potentials, respectively,  see Equations (\ref{phir4}) and (\ref{phiN4}). We will assume, for~simplicity, that the galaxy is cylindrically symmetric and that rotation curves are evaluated at the galactic plane. Thus, according to Equation (\ref{phiN4}), we need to numerically evaluate the integral:
\beq
\psi = \frac{1}{\Lambda} \int_{0}^{2 \pi} d \theta' \int_{0}^{\infty} d \br' \br'  \int_{-\infty}^{\infty} d z'
  \frac{\br}{R} \tilde \rho, \qquad  \tilde \rho = e^{kp (\br)} e^{-\frac{z^2}{\sigma^2}},
\label{psi2}
\enq
in the above:
\beq
\Lambda = \int_{0}^{2 \pi} d \theta' \int_{0}^{\infty} d \br' \br'  \int_{-\infty}^{\infty} d z'  \tilde \rho = 0.721007
\label{lambda3}
\enq
and:
\beq
 \frac{\br}{R} =\frac{1}{\sqrt{1 + \left(\frac{\br'}{\br}\right)^2  -   2 \left(\frac{\br'}{\br}\right) \cos(\theta') + \left(\frac{z'}{\br}\right)^2}}.
\label{Ror}
\enq

\begin{figure}[H]
\centering
\includegraphics[width=0.9\columnwidth]{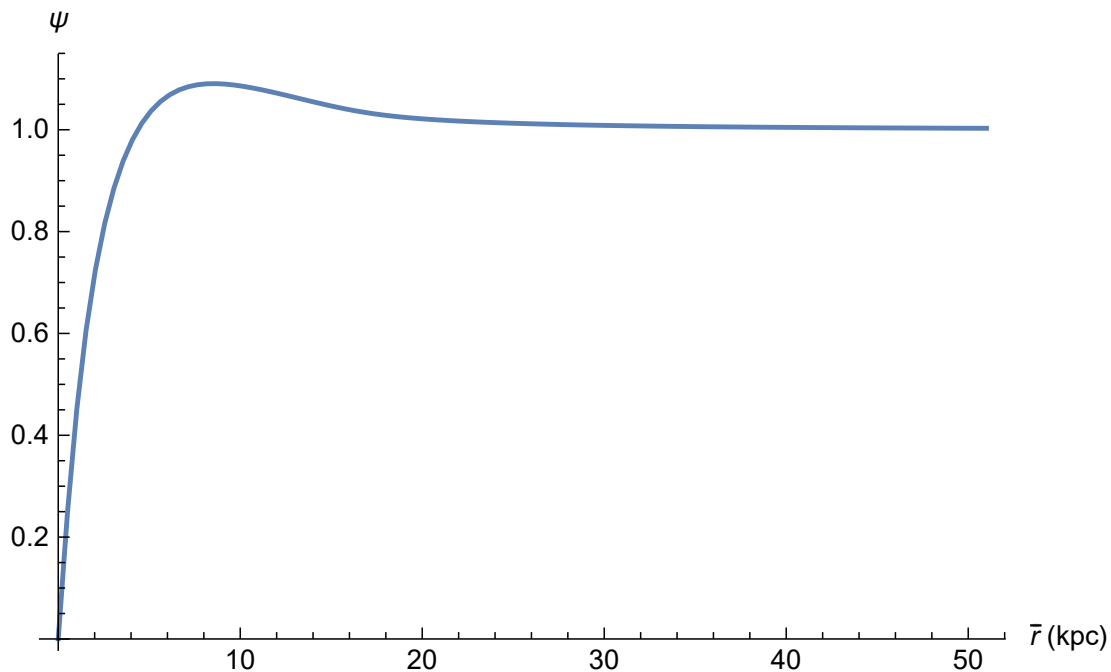}
 \caption{The $\psi$ function of M33, the~function converges to one for large distances as expected from Equation~(\ref{psi}).}
 \label{psip}
\end{figure}

\begin{figure}[H]
\centering
\includegraphics[width=\columnwidth]{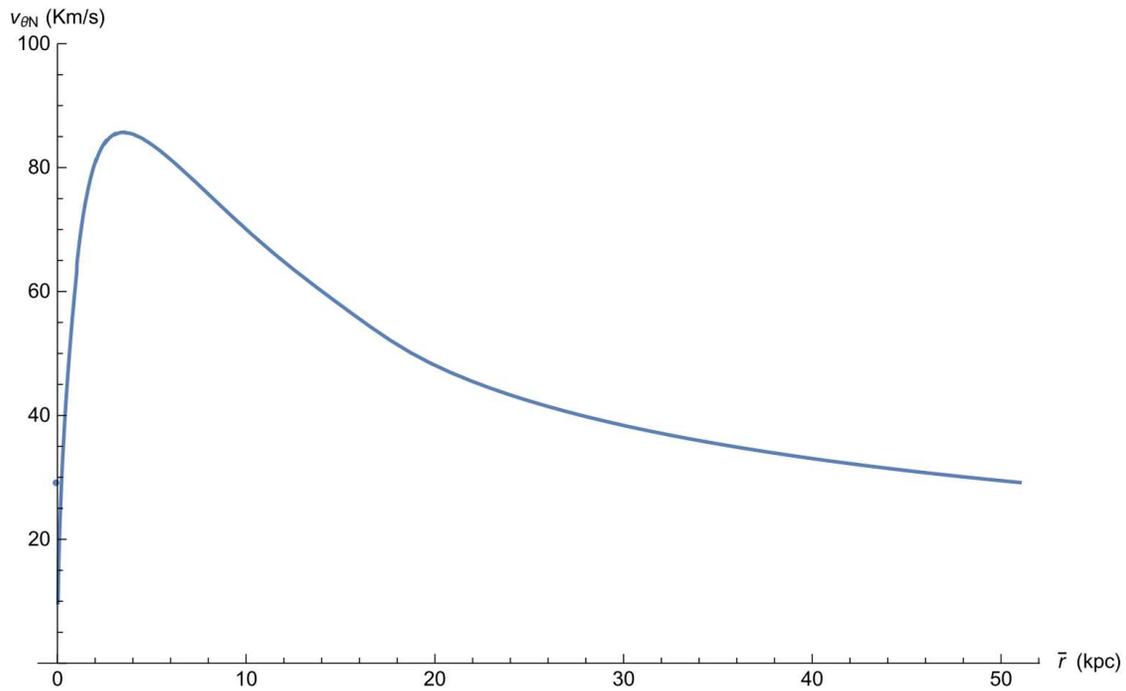}
 \caption{The Newtonian rotation curve of M33; the velocity decreases at large~distances.}
 \label{vtN}
\end{figure}

The result of the numerical evaluation of $\psi(\br)$ is given in Figure~\ref{psip}; we notice that, due to the cylindrical symmetry, it is enough to evaluate the function for the azimuthal angle $\theta=0$ as the result is similar for each azimuthal angle. As~can be easily seen, the~$\psi$ function of M33 converges to one for large distances, as~expected from Equation~(\ref{psi}).
If we ignore the retardation potential contribution Equation~(\ref{Eulersr3}), then Equation~(\ref{phiN4}) will lead to the rotation curve depicted in Figure~\ref{vtN}, which decreases for large distances. In~the above, we assumed a galactic (baryonic) mass of $10^{10}$ solar masses ($2  \times 10^{40}$ kg)  \cite{Corbelli2}.
As the true galactic rotation curve does not decrease, we need to evaluate the retardation potential Equation~(\ref{phir4}) as well. For~this,
we evaluate the $\chi$ function as follows:
\beq
\chi \equiv  \frac{1}{\Lambda} \int_{0}^{2 \pi} d \theta' \int_{0}^{\infty} d \br' \br'  \int_{-\infty}^{\infty} d z'  \frac{R}{r} \tilde \rho
\label{chi2}
\enq

\begin{figure}[H]
\centering
\includegraphics[width=\columnwidth]{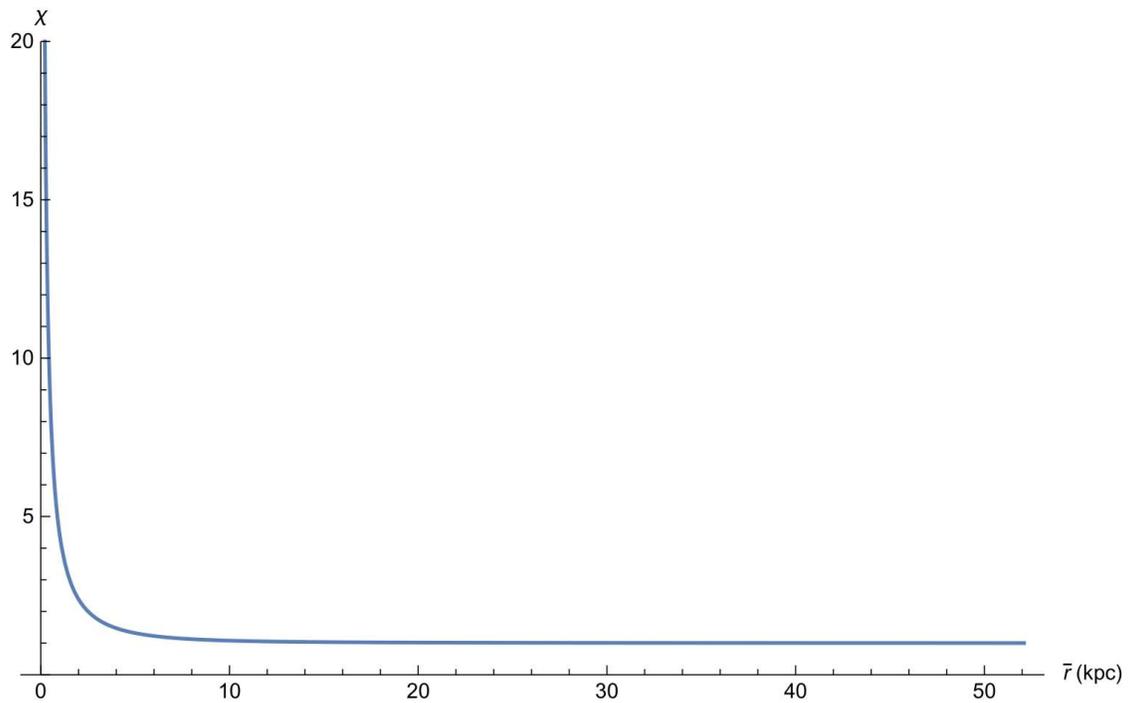}
 \caption{The $\chi$ function of M33; the function converges to one for large distances, as~expected from Equation~(\ref{chi}).}
 \label{chip}
\end{figure}

\begin{figure}[H]
\centering
\includegraphics[width=\columnwidth]{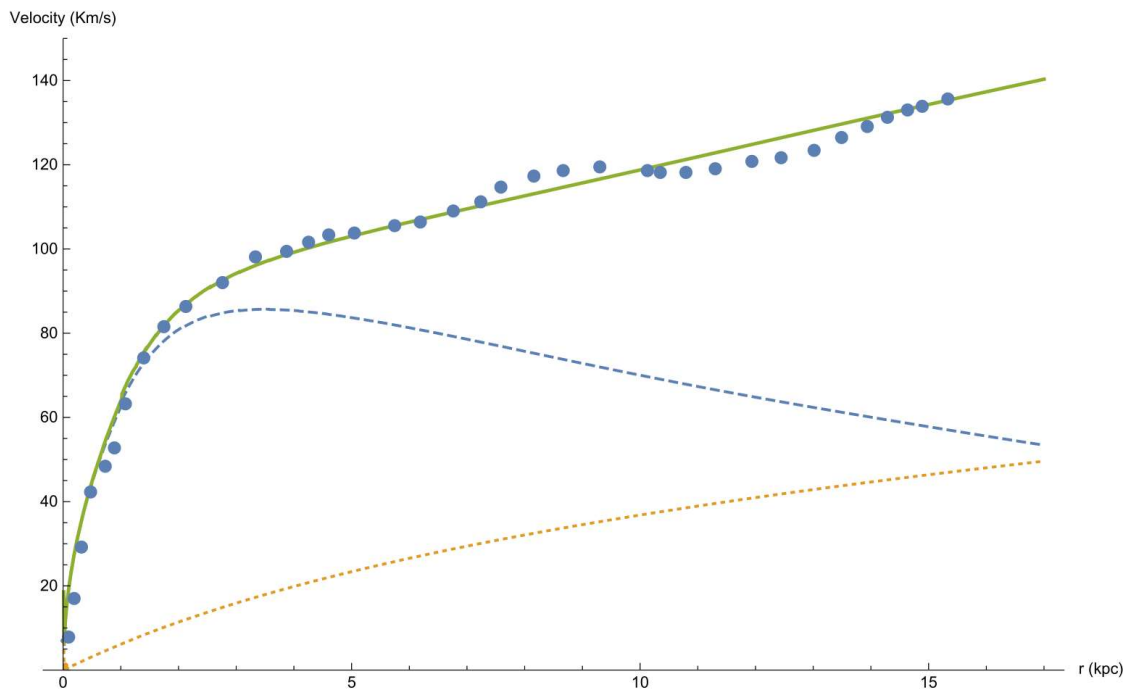}
 \caption{Rotation curve for M33. The~observational points were supplied by Dr. Michal Wagman, a~former PhD student at Ariel University, under~my supervision, using~\cite{Corbelli2}; the full line describes the complete rotation curve, which is the sum of the dotted line, describing the retardation contribution, and~the dashed line, which is the Newtonian~contribution.}
 \label{vcrhoc2}
\end{figure}

The result of the evaluation is given in Figure~\ref{chip}, where it is obvious that the $\chi$ function approaches unity for large distances from the galactic center. The~retardation potential and thus the retardation contribution to the velocity cannot be calculated without knowledge of the retardation distance $R_r$.
However, this can be obtained easily by fitting the observational galactic rotation curve, as~demonstrated in Figure~\ref{vcrhoc2}, which yields
a best fit for $R_r = 4.54$ kpc and a retardation time of $t_r = 14,818.7$ years. Looking at Figure~\ref{vcrhoc2}, this seems quite~reasonable.
We deduce from the above and from Equation~(\ref{deltat}) that the galactic mass second derivative is as~follows:
\beq
|\ddot{M}| = \frac{M}{t_r^2} \simeq 9.12 \times10^{16} \ {\rm kg/s^2}
\label{ddmobse}
\enq

 We underline that this is based only on the {\bf current} estimation of the second derivative of mass; we make no claim  about the past or future values of $\ddot{M}$, nor is there any claim in this section on the value of
$\dot{M}$ at any time or the value of $M$ in the past or the future. It is obvious that such questions involve an understanding of the mass exchange between the galaxy and the intergalactic medium, as~described in Section~\ref{dynmodel}.

\section{A Dynamical~Model}
\label{dynmodel}

\begin{figure}[H]
\centering
\includegraphics[width=0.9\columnwidth]{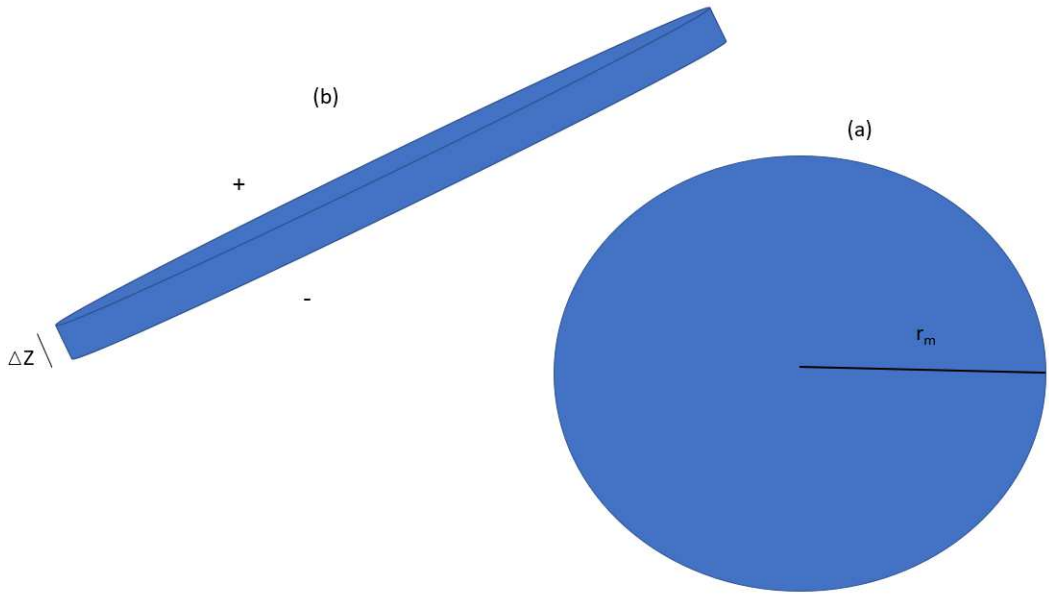}
 \caption{An idealized cylindrical galaxy from different perspectives. (\textbf{a}) From above; (\textbf{b}) tilted edge~perspective.}
 \label{idgalax}
\end{figure}

The mass accumulation model described in the previous section is based on a fitting of the second derivative of the galactic mass to the
galactic rotation curve. It is intuitively obvious that, as~mass is accumulated in the galaxy, it must be depleted in the intergalactic medium.
This is due to the fact that the total mass is conserved; still, it is of interest to see if this intuition is compatible with a model of gas dynamics. For~simplicity, we assume that the gas is a barotropic ideal fluid and its dynamics are described by the Euler and continuity equations as follows:
\beq
\frac{\partial{\rho}}{\partial t} + \vec \nabla \cdot (\rho \vec v ) = 0
\label{masscon}
\enq
\beq
\frac{d \vec v}{d t} \equiv
\frac{\partial \vec v}{\partial t}+(\vec v \cdot \vec \nabla)\vec v  = -\frac{\vec \nabla p (\rho)}{\rho} - \vec \nabla \phi
\label{Euler}
\enq
where the pressure $p (\rho)$ is assumed to be a given function of the density,  $\frac{\partial }{\partial t}$
is a partial temporal derivative, $\vec \nabla$ has its standard meaning in vector analysis and $\frac{d }{d t}$ is the material temporal derivative. We have neglected viscosity terms due to the low gas density. For simplicity, we assume axial symmetry; hence, all variables are independent of $\theta$. Moreover, the~mass influx coming from above and below the galaxy is much more significant compared to the influx coming from the galactic edge. This is due to the large difference in the galaxy surfaces perpendicular to the z axis compared to the area of its edge. The~area of the surface of the galaxy which is perpendicular to the $z$ axis is:
\beq
S_z= S_{z+}+S_{z-} = \pi r_m^2 + \pi r_m^2 = 2 \pi r_m^2
\label{galsz}
\enq
in which $S_z$ is the total surface area of the galaxy perpendicular to the $z$ axis, $S_{z+}$ is the upper area of the surface of the galaxy perpendicular to the $z$ axis, $S_{z-}$ is the lower area of the surface of galaxy perpendicular to the $z$ axis and $r_m$ is the galactic radius
(see Figure~\ref{idgalax}).

The area of the surface of the galactic edge  with thickness $\Delta z$ is:
\beq
S_e= 2 \pi r_m \Delta z.
\label{galse}
\enq

Thus, the~ratio of the surface area is:
\beq
\frac{S_e}{S_z}= \frac{\Delta z}{r_m}.
\label{galsra}
\enq

\begin{figure}[H]
\centering
\includegraphics[width=0.9\columnwidth]{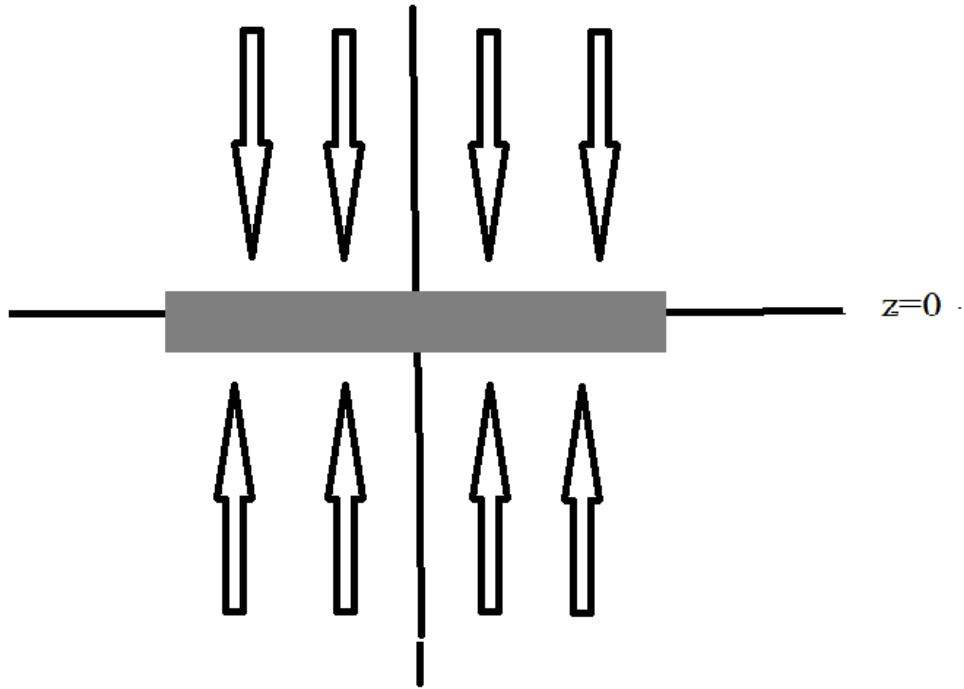}
 \caption{A schematic view of the galactic influx from the~side.}
 \label{influx}
\end{figure}

Typical values of $\Delta z$ are about $0.4$ kilo parsec and $r_m$ is about $17$ kilo parsec (for M33), giving~an area ratio of about $1\%$.
In such circumstances, the~edge mass influx is less important and we  can assume a velocity field of the form:
\beq
\vec v = v_z (\br,z,t) \hat z + v_\theta (\br,z,t) \hat \theta.
\label{velfield}
\enq
$\hat z$ and $\hat \theta$ are unit vectors in the $z$ and $\theta$ directions, respectively.
The influx is described schematically in Figure~\ref{influx}.
In this case, the~continuity Equation~(\ref{masscon}) will take the form:
\beq
\frac{\partial{\rho}}{\partial t} + \frac{\partial{(\rho v_z)}}{\partial z} = 0
\label{masscon2}
\enq

By defining the quantity:
\beq
\gamma \equiv \rho v_z \Rightarrow  \rho = \frac{\gamma}{v_z}
\label{gamma}
\enq
and using the above definition, Equation~(\ref{masscon2}) takes the form:
\beq
\frac{\partial{(\frac{\gamma}{v_z})}}{\partial t} + \frac{\partial{\gamma}}{\partial z} = 0
\label{gammaeq}
\enq

Assuming, for~simplicity, that $v_z$ is stationary and defining the auxiliary variable $t_z$:
\beq
t_z \equiv \int \frac{dz}{v_z}
\label{tz}
\enq
we arrive at the equations:
\beq
\frac{\partial{\gamma}}{\partial t} + \frac{\partial{\gamma}}{\partial t_z} = 0.
\label{gammaeq2}
\enq

This equation can be solved easily, as~follows:
\beq
\gamma (\br,z,t) = f(t-t_z), \qquad  f(-t_z) =  \gamma (\br,z,0) = v_z \rho (\br,z,0)
\label{gamma2}
\enq
for the function $f(x)$, which is fixed by the density initial conditions and the velocity~profile.

Let us now turn our attention to the Euler Equation~(\ref{Euler}); for stationary flows, it takes the form:
\beq
(\vec v \cdot \vec \nabla)\vec v  = -\frac{\vec \nabla p (\rho)}{\rho} - \vec \nabla \phi
\label{Eulers}
\enq

According to Equation~(\ref{velfield}):
\beq
\vec v \cdot \vec \nabla = v_z \frac{\partial}{\partial z} +\frac{v_\theta}{\br} \frac{\partial}{\partial \theta}
\label{vcon}
\enq

Now, by~writing Equation~(\ref{Eulers}) in terms of its components, we arrive at the following equations:
\beq
v_z \frac{\partial v_z }{\partial z}  = -\frac{1}{\rho} \frac{\partial p }{\partial z} - \frac{\partial \phi }{\partial z}
\label{Eulersz}
\enq
\beq
-  \frac{v_\theta^2}{\br}   = -\frac{1}{\rho} \frac{\partial p }{\partial \br} - \frac{\partial \phi }{\partial \br},
 \qquad \left(\frac{\partial \hat{\theta} }{\partial \theta} = - \hat{\br}\right).
\label{Eulersr}
\enq

It is usually assumed that the radial pressure gradients are negligible with respect to the gravitational forces and thus we arrive at the
equation:
\beq
 \frac{v_\theta^2}{\br} \simeq  \frac{\partial \phi }{\partial \br},
\label{Eulersr2}
\enq

As for the $z$-component part of the Euler Equation (\ref{Eulers}), it can be easily written in terms of the specific enthalpy $w(\rho) = \int \frac{dP}{\rho}$ in the form:
\beq
\frac{\partial }{\partial z}  \left(\frac{1}{2} v_z^2 + w(\rho) + \phi \right) = 0 \Rightarrow \frac{1}{2} v_z^2 + w(\rho) + \phi = C(r,t).
\label{Eulersz3}
\enq

We recall that $\rho$ depends on $v_z$ through Equations~(\ref{gamma}) and (\ref{gamma2}):
\beq
\rho(r,z,t) =  \frac{\gamma}{v_z} = \frac{f(t- \int \frac{dz}{v_z})}{v_z}
\label{rho2}
\enq

\begin{figure}[H]
\centering
\includegraphics[width=0.9\columnwidth]{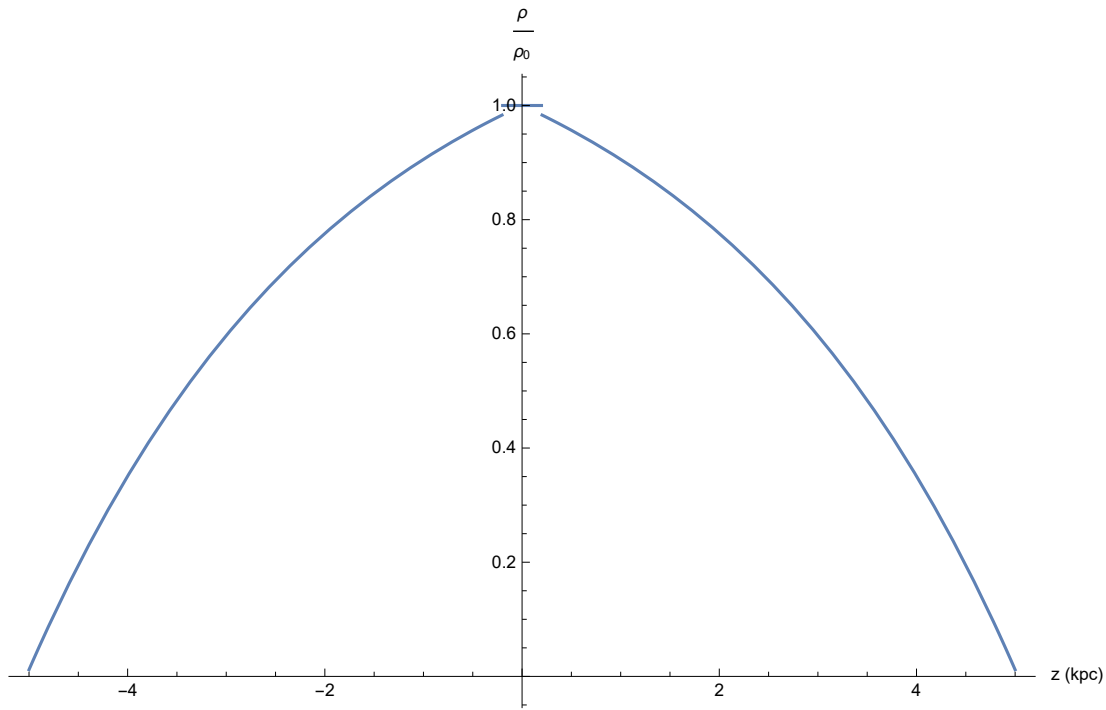}
 \caption{An initial density profile outside the galactic plane, in~which $\rho_0 = \rho_1+\rho_2, \frac{\rho_2}{\rho_1}=-0.2$ and $z_i=5$ (kpc) and $k=0.32 \ \rm (kpc^{-1})$.}
 \label{indenprof}
\end{figure}

As both the specific enthalpy and the gravitational potential are dependent on the density, Equation~(\ref{Eulersz3}) turns into a rather complicated
nonlinear integral equation for $v_z$. However, many~galaxies are flattened structures; hence, it can thus be assumed that the pressure $z$ gradients are significant as one approaches the galactic plane. We will thus assume, for~the sake of simplicity, that~the pressure gradients balance the gravitational pull of the galaxy and thus $v_z$ is just a function of $r$, in~which case the convective derivative of $v_z$ vanishes. The~above assumption holds below and above the galactic plane, but~not at the galactic plane itself. This suggests the following simple model for the velocity $v_z$ (see Figure~\ref{influx}):
\beq
v_z=\left\{
      \begin{array}{cc}
        -|v_z| & z>0 \\
        |v_z| & z<0 \\
      \end{array}
    \right.
\label{vz}
\enq
in which $|v_z|$ is a known function of $\br$. The~velocity field is discontinuous at the galactic plane due to our simplification assumptions, but, of~course, need not be so in reality. We also assume, for~simplicity, that the velocity field $|v_z|$ is constant for $\br<r_m$ and  vanishes for $\br>r_m$. According to Equation~(\ref{gamma2}), the~time-dependent density profile is fixed by the initial density conditions. In~this section, we will deal with the density profile outside the galactic plane and will leave the discussion of the density profile in and near the galactic plane to the previous section. We consider an initial density profile as follows:
\ber
\rho_o(\br,z,0)&=&re(z)\left[\rho_1 (\br) + \rho_2 (\br) e^{k|z|}\right],
 \nonumber \\
  re(z)&=&\left\{           \begin{array}{cc}
                                                                                       1 & |z|<z_i \\
                                                                                       0 & |z| \ge~z_i \\
                                                                                     \end{array}
                                                                                   \right.
\label{rho0}
\enr
in which the rectangular function $re(z)$ keeps the exponential function from diverging. The~density profile is depicted in Figure~\ref{indenprof}.
We assume that $\rho_2$ is negative and thus the density becomes dilute at distances far from the galactic plane.
As $v_z$ is constant both above and below the galactic plane, $t_z=\frac{z}{v_z}$ up to a constant. Therefore, it is easy to deduce from Equation~(\ref{gamma2}) the functional form of $f(\beta)$ {\bf ($\beta=-t_z$)}:
\beq
f(\beta) = v_z re (-v_z \beta) [\rho_1 + \rho_2 e^{k|v_z \beta|}]
\label{fder}
\enq

Hence, according to Equation~(\ref{rho2}), the~time-dependent density function for matter outside the galactic plane is obtained:
\beq
\rho_o(\br,z,t) =  \frac{\gamma}{v_z}=  re (z - v_z t) [\rho_1 (\br)  + \rho_2 (\br) e^{k|z - v_z t|}]
\label{rho3}
\enq

\begin{figure}[H]
\centering
\includegraphics[width=0.9\columnwidth]{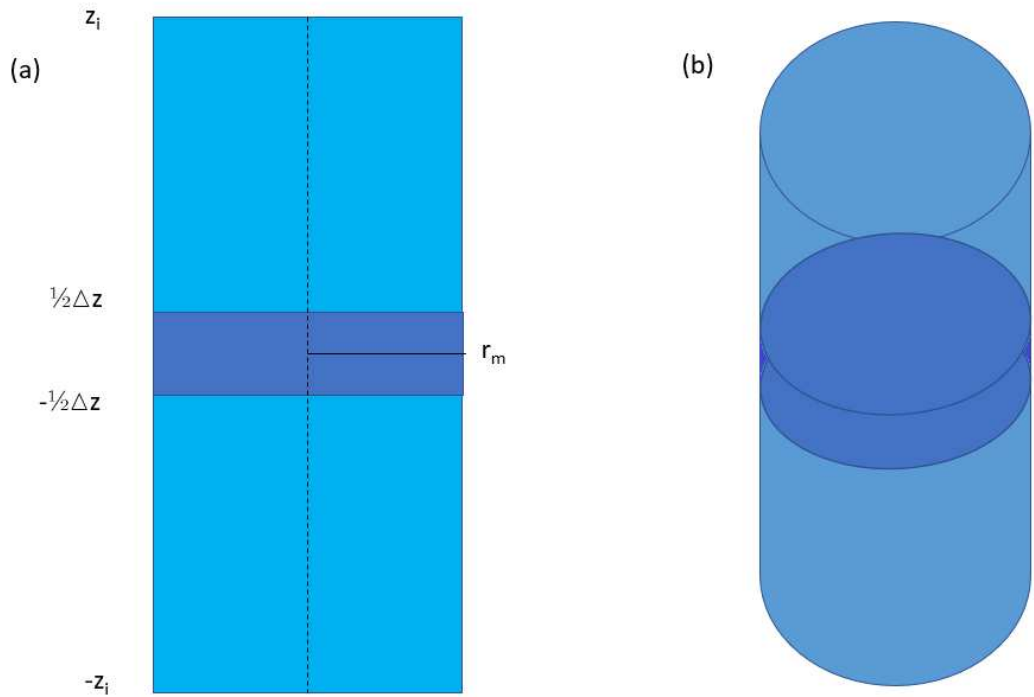}
 \caption{The mass column defined by the galaxy: (\textbf{a}) side view; (\textbf{b}) three-dimensional~view.}
 \label{masscol}
\end{figure}

The density of matter outside the galactic plane will vanish for $t>t_m = \frac{z_i}{|v_z|}$; hence, we will discuss only the duration of $t<t_m$.
Let us look at the mass contained in the cylinder defined by the galaxy  (see Figure~\ref{masscol}) and let us assume that the total mass in that cylinder is $M_T$. Now, the~mass outside the galactic disk will be:
\ber
M_o (t)&=& 2\pi \left[\int_{-z_i}^{-\frac{1}{2} \Delta z} dz \int_{0}^{r_m} d \br \br \rho_o(\br,z,t) \right.
\nonumber \\
&+&
 \left. \int_{\frac{1}{2} \Delta z}^{z_i} dz \int_{0}^{r_m} d \br \br \rho_o(\br,z,t) \right]
\label{Mo}
\enr

Hence, the~mass in the galactic disk is:
\beq
M(t) = M_T - M_o (t)
\label{Mg}
\enq

Moreover, the~galactic mass derivatives are:
\beq
\dot{M}(t) =  - \dot{M_o}(t), \qquad \ddot{M}(t) =  - \ddot{M_o}(t)
\label{Mgd}
\enq

By inserting Equation~(\ref{rho3}) into Equation~(\ref{Mo}), we may calculate $M_o (t)$:
\ber
M_o (t) &=& 2 \left[\lambda_1 \left(z_i-|v_z|t-\hdz\right) \right.
\nonumber \\
&+& \left. \frac{\lambda_2}{k} \left(e^{k z_i}- e^{k (|v_z|t +\hdz)}\right)\right]
\label{Mo2}
\enr
in which:
\beq
\lambda_1 \equiv 2 \pi \int_{0}^{r_m} d \br \br \rho_1 (\br), \qquad \lambda_2 \equiv 2 \pi \int_{0}^{r_m} d \br \br \rho_2 (\br).
\label{lam}
\enq

Then, calculating the second derivative of $M_o (t)$ and using Equation~(\ref{Mgd}) leads to the result:
\beq
 \ddot{M}(t) =  - \ddot{M_o}(t) = 2 k |v_z|^2 e^{\hdz k} \lambda_2  e^{k |v_z|t}.
\label{Mgd2}
\enq

We then denote:
\beq
 \alpha \equiv k |v_z|, \qquad \tau \equiv \frac{1}{\alpha}, \qquad \ddot{M}(0) = 2 k |v_z|^2 e^{\hdz k} \lambda_2.
\label{Mgd3}
\enq

Thus:
\beq
\ddot{M}(t) = \ddot{M}(0) e^{\frac{t}{\tau}}.
\label{ddMt4}
\enq

 It should be stressed that the current approach does not require that the velocity $|v_z|$ is high; in fact, the~vast majority of galactic bodies (stars, gas) are substantially subluminal—in other words, the~ratio of $\frac{|v_z|}{c} \ll 1$. The~typical velocities in galaxies are ~100 km/s, which makes this ratio $0.001$ or smaller.
However, we stress again the fact that every gravitational system, even if it is made of subluminal bodies, has a finite retardation distance, beyond~which the retardation effect cannot be neglected.  As~demonstrated above, a~galaxy exchanges mass with its environment. This means that a galaxy has a finite retardation distance. The~question is thus quantitative: how large is the retardation distance? For the M33 galaxy, the~velocity curve indicates that retardation effects cannot be neglected beyond a certain distance, which is calculated in Section~\ref{RoCu} to be roughly 14,000 light years; similar analyses for other galaxies of different types have shown similar results~\cite{Wagman}. We demonstrate here, using a detailed model, that this does not require a high velocity of gas or stars, in~or out of the galaxy, and~is perfectly consistent with the current observational knowledge of galactic and extragalactic material content and dynamics.
Equation~(\ref{Mgd3}) means that we must have $\lambda_2<0$ according to Equation~(\ref{Fr3}) in order to assure an attractive force.
Next, we calculate $\dot{M}(t)$; using Equations~(\ref{Mgd}) and (\ref{Mo2}), we obtain:
\beq
 \dot{M}(t) =  - \dot{M_o}(t) = 2 |v_z| \lambda_1 +  2  |v_z| e^{\hdz k} \lambda_2  e^{k |v_z|t}.
\label{Mgd4}
\enq

Hence:
\beq
 \dot{M}(0) =   2 |v_z| \lambda_1 +  2  |v_z| e^{\hdz k} \lambda_2  =  2 |v_z| \lambda_1 +\frac{\ddot{M}(0)}{\alpha}
\label{Mgd5}
\enq

Thus, $\lambda_1$ is:
\beq
 \lambda_1 = \frac{1}{ 2 |v_z|} \left[\dot{M}(0)  - \frac{\ddot{M}(0)}{\alpha} \right]
\label{Mgd6}
\enq

Inserting Equations~(\ref{Mgd6}) and (\ref{Mgd3}) into Equation~(\ref{Mgd4}) leads  to:
\beq
\dot{M}(t) = \dot{M}(0) + \tau \ddot{M}(0) \left(e^{\frac{t}{\tau}} - 1\right)
= \dot{M}(0) -\tau |\ddot{M}(0)| \left(e^{\frac{t}{\tau}} - 1\right).
\label{dMt4}
\enq

Finally, by~combining Equations~(\ref{Mg}), (\ref{Mo2}), (\ref{Mgd3}) and  (\ref{Mgd6}) and noticing that:
\beq
{M}(0) =  M_T  - \frac{z_i-\hdz}{  |v_z|} \left(\dot{M}(0)  - \frac{\ddot{M}(0)}{\alpha} \right)
- \frac{\ddot M (0)}{\alpha^2} \left( e^{k(z_i -\hdz)} - 1 \right)
\label{Mgd7}
\enq
we arrive at:
\beq
M(t) = M(0)+ \left(\dot{M}(0)-\tau \ddot{M}(0)\right) t + \tau^2 \ddot{M}(0) \left(e^{\frac{t}{\tau}} - 1\right), \quad \tau>0.
\label{Mt4}
\enq

\section{The Mass~Formula}
\label{Massform}

A naive observer may speculate that $\ddot{M}$, given in Equation~(\ref{ddmobse}), is constant over time; however, this  is contradicted both
by the dynamical model as presented in  Equation~(\ref{Mt4}) and by the following simple~argument.

Let us assume that $\ddot{M}$ is constant. This leads to:
\beq
\dot{M}(t)= \dot{M}(0) + t \ddot{M} = \dot{M}(0) - t |\ddot{M}|
\label{dMt}
\enq

Let us assume that there is no significant mass accumulation at the current time $T$, that is, $\dot{M}(T)=0$. In~this case:
\beq
 \dot{M}(0) = T |\ddot{M}|
\label{dMt0}
\enq

Assuming that the age of the galaxy is about the age of the universe:  $T \simeq 13 \times10^9 \ {\rm years} = 4.1 \times 10^{17}  {\rm s}$, the~initial mass accumulation rate is about $\dot{M}(0) \simeq 3.74 \times 10^{34}{\rm~kg \ s^{-1}}$. By~integrating Equation~(\ref{dMt}), we~arrive
at an equation for the mass of the galaxy at any time:
\beq
M(t) = M(0) + \dot{M}(0) t + \frac{1}{2} \ddot{M} t^2
\label{Mt2}
\enq

Assuming that, initially, there was no significant amount of mass $M(0)=0$ and taking into account Equation~(\ref{dMt0}) (as for Equation (\ref{dMt0}), it should be once again recalled that the second derivative  of $M$ by $t$ is negative) we arrive at:

\beq
M(t) =\ddot{M} t (\frac{1}{2} t - T) \Rightarrow  M(T) =-\frac{1}{2} \ddot{M} T^2= \frac{1}{2} |\ddot{M}| T^2
\label{Mt3}
\enq

By plugging in the mass accumulation decrease rate, we arrive at $ M(T) \simeq 7.66\times 10^{51}$ kg, which is clearly 11 orders of magnitude greater
than the known mass of the galaxy. On~the other hand, if~we assume that $M(T)=M$ is the current mass of the galaxy, this will lead
to $|\ddot{M}| \simeq  2.37\times 0^{5} $  kg ${\rm {s}^{-2}}$, which~is 11 orders of magnitude less than what is needed to explain the current galactic
rotation curves. This~indicates that $|\ddot{M}| $ has increased considerably over time due to the depletion of the surrounding gas and that
retardation forces are less significant in the early stages of galactic formation. Indeed, in~young galaxies, the~effect of retardation seems insignificant~\cite{Dokkum}, while, for~older galaxies, it seems like somebody is pressing hard on the brakes of mass accumulation. The~changes, by~many orders of magnitude, in~$|\ddot{M}|$ suggest exponential growth in agreement with Equation~(\ref{Mt4}).

If we assume $\dot{M}(t) > 0$, it follows from Equation~(\ref{dMt4}) that there is a maximal duration for the above
model to be valid:
\beq
t_{max1} = \tau \ln \left(\frac{\dot{M}(0)}{\tau |\ddot{M}(0)|} + 1 \right).
\label{dMt5}
\enq

This is the time it takes the intergalactic gas to deplete (see Section~\ref{dynmodel}) and thus mass accumulation stops.
We shall now partition the mass into a linear and exponential growing part as follows:
\beq
M(t) =  M_l (t) + M_e (t) = M_l (t) - |M_e (t)| > 0 \Rightarrow M_l (t) > |M_e (t)|.
\label{Mt5a}
\enq
\ber
M_l (t) & = &  M(0) + \tau^2 |\ddot{M}(0)| + \left(\dot{M}(0)+ \tau |\ddot{M}(0)|\right) t > 0,
\nonumber \\
M_e(t) & = &  \tau^2 \ddot{M}(t) =  - \tau^2 |\ddot{M}(0)| e^{\frac{t}{\tau}} < 0.
\label{Mt6}
\enr

By dividing Equation~(\ref{Mt5a}) by $|\ddot{M}(t)|$  and using the definition of Equation~(\ref{deltat}), we have:
\beq
t_r^2  = \frac{M_l (t)}{|\ddot{M}(t)|} - \tau^2.
\label{Mt7}
\enq

We define the slow retardation time as:
\beq
t_{rs} (t) \equiv  \sqrt{\frac{M_l (t)}{|\ddot{M}(0)|}};
\label{Mt8}
\enq
the square of this retardation time is non-linearly dependent on time. Hence: 
\beq
t_r^2  =t_{rs}^2 e^{-\frac{t}{\tau}} - \tau^2
\label{Mt9}
\enq

It is obvious that $t_{rs} > \tau$ and it is also obvious that for any $t<t_{max1}$, then $t_{rs} e^{-\frac{t}{2\tau}} > \tau$ because
the total mass is not reduced, only  mass accumulation slows down (in other words, we assume that the galaxy does not eject matter, thus reducing its mass). \Ern{Mt9} shows that there is no simple relation between the retardation time of the galaxy and the typical time $\tau$ associated with the second derivative of~mass.

We conclude this section by deriving constraints from the assumption that the apparent luminosity and rotation curve have not changed in a detectable way since observations of the M33 galaxy have begun. If~the mass to light ratio is assumed to be constant, this means that the mass of the galaxy has not changed significantly during the said period. Furthermore, the~retardation time, which has a significant effect on the galactic rotation curve, is a square root of the ratio of the galactic mass by the second derivative of the same mass. This means that if the retardation time (and retardation distance) have not changed considerably during the duration of observations, then the second derivative must also be approximately constant during that period. The~M33 Triangulum Galaxy was probably discovered by the Italian astronomer Giovanni Battista Hodierna before 1654. In~his work, De systemate orbis cometici; deque admirandis coeli caracteribus (``About the systematics of the cometary orbit, and~about the admirable objects of the sky''), he listed it as a cloud-like nebulosity or obscuration and gave the cryptic description, ``near the Triangle hinc inde''. This is in reference to the constellation of Triangulum as a pair of triangles. The~magnitude of the object matches M33, so it is most likely a reference to the Triangulum galaxy~\cite{Fodera}. This observation took place 366 years ago; however, there~was no measurement of the apparent or real luminosity of M33 and of course no measurement of Doppler shifts were available at that time. It was only Hubble that established the distance to this object and thus declared it an independent galaxy far away from the Milky Way~\cite{Van}, thus enabling the calculation of the real luminosity and mass of M33. Doppler shifts of spectral lines were obtained even later, thus~enabling the deduction of rotation curves. In~what follows, we shall assume an observation duration of:
\beq
\delta t = {\rm 400 \ years}
\label{Mt10}
\enq

As $\ddot{M}(t)$ is approximately constant during the said duration, it follows that:
\beq
|\frac{d^3 M(t)}{dt^3}\delta t| \ll |\ddot{M}(t)| .
\label{ddMt11}
\enq

Taking into account Equation~(\ref{ddMt4}) gives a lower bound to $\tau$:
\beq
\delta t \ll \tau.
\label{ddMt12}
\enq

This is not a strong bound as $\tau$ can be anything from 100,000 years to ten times the age of the universe or even larger. 
Similarly,
the fact that the mass of the galaxy has not changed considerably during the observation duration indicates that:
 \beq
|\dot{M}(t)| \delta t| \ll |M(t)|  \Rightarrow  \dot{M}(t) \ll \dot{M}_{max} \equiv \frac{M(t)}{\delta t}
\label{Mt11}
\enq

The upper bound on the mass derivative is thus about $\dot{M}_{max} \simeq 1.6\times 10^{30} $ kg/s, in~which we have taken the baryonic mass of the M33 galaxy to be about $2 \times 10^{40}$ kg~\cite{Corbelli2}. Using Equation~(\ref{dMt4}), we arrive with an upper bound to $\dot{M}(0)$:
\beq
\dot{M}(0) \ll \dot{M}_{max} + \tau  \left(|\ddot{M}(t)| - |\ddot{M}(0)| \right).
\label{dMt4b}
\enq

Notice, however, that without a handle on $\tau$ and $|\ddot{M}(0)|$, this does not yield any definite~information.

\section{Conclusions}

The need to satisfy the Lorentz symmetry group prevents the weak field approximation of GR from allowing action at
distance potentials and thus only retarded solutions are allowed. Retardation is manifested more strongly when large distances
and large second derivatives are involved. It should be stressed that the current approach does not require that velocities, $v$ are high; in fact, the~vast majority of galactic bodies (stars, gas) are substantially subluminal—in other words, the~ratio of $\frac{v}{c} \ll 1$. The~typical velocities in galaxies are $100 {\rm \frac{km}{s}}$ (see Figure~\ref{vcrhoc2}), which makes this ratio $0.001$ or smaller.
However, one should consider the fact that every gravitational system, even if it is made of subluminal bodies, has a retardation distance, beyond~which the retardation effect cannot be neglected.  Every natural system, such as a star or a galaxy and even a galactic cluster, exchanges mass with its environment. For~example, the~sun loses mass through solar wind and galaxies accrete gas from the intergalactic medium. This means that all natural gravitational systems have a finite retardation distance. The~question is thus quantitative: how large is the retardation distance? The change in the mass of the sun is quite small and thus the retardation distance of the solar system is huge, allowing~us to neglect retardation effects within the solar system. However, for~the M33 galaxy, the~velocity curve indicates that the retardation effects cannot be neglected beyond a certain distance, which was calculated in Section~\ref{RoCu} to be roughly $R_r = 4.54$ kpc; similar analyses for other galaxies of different types have shown similar results~\cite{Wagman,Wagman2}. We demonstrated, using a detailed model, in~Section~\ref{dynmodel},  that~this does not require a high velocity of gas or stars in or out of the galaxy and is perfectly consistent with the current observational knowledge of galactic and extragalactic material content and~dynamics.

We point out that, if~the mass outside the galaxy is still abundant (or totally consumed), $\ddot{M} \simeq 0$
and retardation force should vanish. This was reported~\cite{Dokkum} for the galaxy~NGC1052-DF2.

We note that the same terms in the gravitation equation that are responsible for the gravitational radiation recently discovered are also responsible for the rotation curves of galaxies. The~expansion given in Equation~(\ref{phir}), being a Taylor series expansion up to the second order, is only valid for limited~radii:
\beq
R < c \ T_{max} \equiv R_{max}
\label{Rmax}
\enq

This means that current expansion is related to the near field case; this is acceptable since the extension of the rotation curve in galaxies
is the same order of magnitude as the size of the galaxy itself. An~opposite case in which the size of the object is much smaller than the distance to the observer will result in a different approximation to Equation (\ref{bhint}), leading to the famous quadruple equation of gravitational radiation, as~predicted by Einstein~\cite{Einstein2} and verified indirectly in 1993 by Russell A. Hulse and Joseph H. Taylor, Jr., for~which they received the Nobel Prize in Physics. The~discovery and observation of the Hulse–Taylor binary pulsar offered the first indirect evidence of the existence of gravitational waves~\cite{Taylor}. On~11 February 2016, the~LIGO 
and Virgo Scientific Collaboration announced that they had made the first direct observation of gravitational waves. The~observation was made five months earlier, on~14 September 2015, using Advanced LIGO detectors. The~gravitational waves originated from the merging of a binary black hole system~\cite{Castelvecchi}. Thus, the~current paper involves a near-field application of gravitational radiation while previous works discuss far- field~results.

We regret that direct measurement of  the second temporal derivative of the galactic mass is not available. What is available is the remarkable fit between the retardation model and the galactic rotation curve, as~can be seen in Figure~\ref{vcrhoc2}, which constitutes indirect evidence of the galactic mass second derivative. The~reader is reminded that competing theories like dark matter do not supply any observational evidence either. Despite the work of thousands of people and the investment of billions of dollars, there is still no evidence of dark matter. Occam's razor dictates that when two theories compete, the~one that makes less assumptions has the upper hand. In~the case of retardation theory, only baryonic matter and a large second temporal derivative of mass are~assumed.

Problems that inflict dark matter theory, such as the cuspy halo problem (also known as the core-cusp problem) refers to a discrepancy between the inferred dark matter density profiles of low-mass galaxies and the density profiles predicted by cosmological N-body simulations. Nearly~all simulations form dark matter halos, which have ``cuspy'' dark matter distributions, with~density increasing steeply at small radii, while the rotation curves of most observed dwarf galaxies suggest that they have flat central dark matter density profiles (``cores''). This problem does not occur in the retardation model which denies the existence of dark matter. One cannot discuss flat or sharp profiles of dark matter if dark matter does not exist. The~inherent problems with dark matter's dynamics further strengthen the claim of this work that dark matter does not exist and the rotation curve characteristics attributed to dark matter should be attributed to~retardation.

How can one reach an erroneous conclusion regarding the existence of ``dark matter''? By ignoring retardation effects and assuming that radial velocities are a result of some mysterious substance. We~obtain for a spherically symmetric distribution~\cite{Binney}:
\beq
-\frac{v_c^2}{r} \hat r = \vec F_d = - \frac{G M_d(r)}{r^2} \hat r
\label{Fd}
\enq
where $v_c$ is the speed of a particle of unchanging radius $r$ and $M_d(r)$ is the dark matter within the radius $r$.
Comparing Equations~(\ref{Fd}) and (\ref{Fr3}), we deduce that the "dark matter" mass can be calculated as follows:
\beq
M_d(r) =   \frac{r^2 |\ddot{M}|}{2  c^2}
\label{Fr4}
\enq

Then, since:
\beq
M_d(r) =  4 \pi \int_{0}^{r} r'^2 \rho_d (r') dr', \qquad \frac{d M_d(r) }{dr} = 4 \pi r^2 \rho_d (r)
\label{Md}
\enq
it follows that:
\beq
\rho_d (r) = \frac{|\ddot{M}|}{4 \pi c^2 r}
\label{rhod}
\enq

This is consistent with the observational data of~\cite{Corbelli} who concluded that the ``dark matter'' density
decreases as $r^{-1.3}$ for~M33.

An additional method to explaining galactic rotation curves is to postulate that either the laws of dynamics or
the laws of gravitation (GR) should be changed. This is the case in an approach championed by Milgrom {(modifying the laws of gravity (MOND)—modified Newtonian dynamics)~\cite{Mond}. In~this approach, the~classical law  of gravity is modified:
\beq
 \vec F_M = - \frac{G M}{\mu(\frac{a}{a_0}) r^2} \hat r
\label{FM}
\enq

In the above $\mu$, is the interpolation function that should be one for $a_0\ll a$. Let us assume:
\beq
  \mu(x) = \frac{x}{\sqrt{1+x^2}} \quad \Rightarrow \quad  \mu(\frac{a}{a_0})=\frac{1}{\sqrt{1+(\frac{a_0}{a})^2}}
\label{mu}
\enq
if $a_0\gg a$, $\mu \simeq \frac{a}{a_0}$. A~particle revolving in a unchanging radius will have a centrifugal acceleration of $a=\frac{v^2}{r}$ and thus:
\beq
 \vec F_M = - \frac{G M a_0}{v^2 \ r} \hat r
\label{FM2}
\enq

For the constant $v$ at a far away distance, this expression is similar to the retardation force
and~thus:
 \beq
 |\ddot{M}| = \frac{2 M a_0 c^2}{v^2 \ r}.
\label{ddM}
\enq

Milgrom found $a_0 = 1.2 \times  10^{-10}{\rm m s^{-2}}$ to be most fitting to the data. The~velocity at $15.33 \ {\rm kpc}$ from the center of the galaxy is 135,640 $\ {\rm m s^{-1}}$. We thus obtain $|\ddot{M}| \simeq  4.94 \times 10^{16} \ {\rm kg s^{-2}}$ and a retardation time~of:
\beq
t_r = \sqrt{\frac{M}{|\ddot{M}|}} \simeq 6.35 \ 10^{11} \ {\rm s}
\label{deltat2}
\enq

 This amounts to a typical accumulation acceleration time scale of $t_r\simeq $20,129 years and a retardation distance of:
\beq
R_r = c t_r\simeq 20,129 {\rm \ light \ years},
\label{Rr2}
\enq
which seems reasonable according to our fitting estimates. Hence, despite the~fact that modified Newtonian dynamics theory is phenomenological and is in contradiction to general relativity it can still serve as tool for estimating retardation theory~quantities.

We have thus shown that ``dark matter'' and ``MOND'' effects can be explained in the framework of
standard GR as effects due to retardation, without~assuming any exotic matter or modifications of the theory of~gravity.

To conclude this section, we would like to mention the remarkable theory of conformal gravity put forward by Mannheim~\cite{Mannheim1,Mannheim2}.
The current retardation approach leads to a Newtonian potential plus a linear potential. Such potential types can also be derived from
the completely different theoretical considerations of conformal gravity. Indeed, on~purely phenomenological grounds, such fits have essentially already been published in the literature. While those fits looked very good, they had to treat the coefficient of the linear potential as a variable that changed from galaxy to galaxy; this element is not in line with conformal gravity in
which the coefficient of the linear potential is a new universal constant of nature. This can be explained much more easily
in the framework of retardation theory, in~which  a variable $\ddot{M}/M$ seems reasonable depending on the dynamical conditions of various galaxies. Indeed, $\ddot{M}/M$ has a theoretical basis, while a pure phenomenological approach does not. As~far as we understand the work of Mannheim~\cite{Mannheim0,Mannheim1,Mannheim2}, it is related to conformal gravity, which is different from GR, and~thus has to justify other results of GR (Big Bang Cosmology, etc.). We also underline that retardation theory does not contradict conformal gravity and, in~reality, both effects may exist, although~Occam's razor forbids us to add new universal constants if the existing ones suffice to explain~observations.

Retardation theory's approach is minimalistic (in the sense that it satisfies the Occam's razor  rule), does not affect observations that are beyond the near-field regime and, therefore, does not clash with GR theory and its observations (nor with Newtonian theory, as~the retardation effect is negligible for small distances). We underline that a perfect fit to the rotation curve is achieved with a single parameter and we do not  adjust the mass to light ratio in order to improve our fit as other authors~do.

Retardation effects in  electromagnetic theory were discussed in~\cite{Tuval,YahalomT,Yahalom3}.

Finally the paper does not discuss dark matter in a cosmological context or in a gravitational lensing scenario; this is left for future works.
Another important application is the study of galaxy clusters. This issue led Zwicky \cite{zwicky} to suggest dark matter in the first place.

\authorcontributions{This paper has a single author, which has done all the work presented.}
 
\funding{This research received no external~funding.}
 
\acknowledgments{The author wishes to thank his former student,  Michal Wagman, for~supplying the data points for the rotation curve of the M33 galaxy.
This work is a result of more than twenty five years of thinking (discontinuously) on the dark matter problem, which was first suggested to
me by the late   Jacob Bekenstein during my stay at the Hebrew University of Jerusalem. The~current retardation solution arose from my discussions with the late  Donald Lynden-Bell of Cambridge University, and~the late Miron Tuval. Other people with which I discussed this work and offered important feedback are   Lawrence Horwitz of Tel-Aviv University and  Marcelo Shiffer of Ariel University.
Recently, this work benefitted from discussions with  James Peebles, Neta Bachall and Sam Cohen, all from Princeton University.
Special thanks are due to   Jiri Bicak for inviting me to present the theory at Charles University in Prague. I would like to thank
Philip Mannheim and   James Obrien
for our discussions during the recent IARD 
meetings and for supplying some relevant data. I have benefited from discussions with a very long list of distinguished scientists and ask their forgiveness for not mentioning them~all. }

\conflictsofinterest{The author declares no conflict of~interest.}

\reftitle{References}

\publishersnote{MDPI stays neutral with regard to jurisdictional claims in published maps and institutional affiliations.}
\end{document}